\documentclass[twocolumn,showpacs,preprintnumbers,amsmath,amssymb]{revtex4}

\usepackage{graphicx}
\usepackage{dcolumn}
\usepackage{bm}
\graphicspath{{./}{./figs/}}

\begin{document}

\title{\bf Synchronization of fractional order chaotic systems \footnote{Phys. Rev. E, Vol. 68, 067203, 2003}}
\author{Chunguang Li$^1$, Xiaofeng Liao$^{1,2}$, Juebang Yu$^1$}
%\email{cgli@uestc.edu.cn}
\affiliation{$^1$Institute of Electronic Systems, School of
Electronic Engineering,\\ University of Electronic Science and
Technology of China,\\ Chengdu, Sichuan, 610054, P. R. China.\\
$^2$School of Computer Science and Technology, Chongqing University,\\
Chongqing, 400044, P. R. China.}
\date{}
\begin{abstract}
The chaotic dynamics of fractional order systems begin to attract
much attentions in recent years. In this brief report, we study
the master-slave synchronization of fractional order chaotic
systems. It is shown that fractional order chaotic systems can
also be synchronized.
\end{abstract}
\maketitle

Fractional calculus is a 300-year-old topic. Although it has a
long mathematical history, the applications of fractional calculus
to physics and engineering are just a recent focus of interest [1,
2]. Many systems are known to display fractional order dynamics,
such as viscoelastic systems [3-5], dielectric polarization [6],
electrode-electrolyte polarization [7] and electromagnetic waves
[8]. More recently, many authors begin to investigate the chaotic
dynamics of fractional order dynamical systems [9-17]. In [9], it
has been shown that the fractional order Chua's system of order as
low as 2.7 can produce a chaotic attractor. In [10], it has been
shown that nonautonomous Duffing systems of order less than 2 can
still behave in a chaotic manner. In [11], chaotic behaviors of
the fractional order ``jerk" model was studied, in which chaotic
attractor was obtained with system orders as low as 2.1, and in
[12] the chaos control of this fractional order chaotic system was
reported. In [13], chaotic behavior of the fractional order Lorenz
system was studied, but unfortunately, the results presented in
this paper are not correct. In [14] and [15], bifurcation and
chaotic dynamics of the fractional order cellular neural networks
were studied. In [16], chaos and hyperchaos in the fractional
order R\"ossler equations were studied, in which we showed that
chaos can exists in the fractional order R\"ossler equation with
order as low as 2.4, and hyperchaos exists in the fractional order
R\"ossler hyperchaos equation with order as low as 3.8. In [17],
we have studied the chaotic behavior and its control in the
fractional order Chen system. In [18], the author present a broad
review of existing models of fractional kinetics and their
connection to dynamical models, phase space topology, and other
characteristics of chaos.

On the other hand, synchronization of chaotic systems has
attracted much attentions [19] since the seminal paper by Pecora
and Carroll [20]. In this brief report, we study the
synchronization of fractional order chaotic systems. The analysis
of fractional order systems is by no means trivial. So, we will
numerically investigate this topic here.

There are many definitions of fractional derivatives [1]. Perhaps
the best known one is the Riemann-Liouville definition, which is
given by
\begin{equation}
\frac{d^\alpha f(t)}{dt^\alpha} = \frac{1} {\Gamma(n-\alpha)}
\frac{d^n}{dt^n} \int_0^t
\frac{f(\tau)}{(t-\tau)^{\alpha-n+1}}d\tau
\end{equation}
where $\Gamma(\cdot)$ is the gamma function and $n-1\leq \alpha
<n$. The geometric and physical interpretation of the fractional
derivatives was given in [21]. Upon considering the initial
conditions to be zero, the Laplace transform of the
Riemann-Liouville fractional derivative is $
L\left\{\frac{d^\alpha f(t)}{dt^\alpha}\right\}=s^\alpha
L\{f(t)\}$. So, the fractional integral operator of order
``$\alpha$" can be represented by the transfer function
$F(s)=\frac{1}{s^\alpha}$.

The standard definition of the fractional differintegral do not
allow direct implementation of the fractional operators in
time-domain simulations. An efficient method to circumvent this
problem is to approximate the fractional operators by using the
standard integer order operators. In the following simulations, we
will use the approximation method proposed in [22], which was also
adopted in [9, 11, 14, 15, 16, 17]. In Table 1 of [9], the authors
gave approximations for $1/s^q$ with $q=0.1-0.9$ in steps 0.1 with
errors of approximately 2dB. We will use these approximations in
our following simulations.

Consider the master-slave synchronization scheme of two autonomous
$n$-dimensional fractional order chaotic systems
\begin{equation}
\begin{array}{rl}
M&: \frac{d^\alpha x}{dt^\alpha}=f(x)\\
S&: \frac{d^\alpha y}{dt^\alpha}=f(y)+c\Gamma(x-y)
\end{array}
\end{equation}
with the master system $M$ and the slave system $S$. Where
$\alpha>0$ is the fractional order, with which the individual
dynamical systems are chaotic, $c>0$ is the coupling strength, and
$\Gamma\in R^{n\times n}$ is a constant 0 - 1 matrix linking the
coupling variables. For simplicity, we assume
$\Gamma=\mbox{diag}(r_1,r_2,\cdots,r_n)$ is a diagonal matrix. If
there is a coupling between the $i$th state variable of the two
coupled chaotic systems, then $r_i=1$; otherwise, $r_i=0$. Define
the error signal as $e=x-y$, the aim of the synchronization scheme
is to design the coupling strength such that $\|e(t)\|\rightarrow
0$ as $t\rightarrow \infty$. This scheme is similar to the
master-slave synchronization of classical integer-order chaotic
systems.

Next, we numerically study the synchronization of fractional order
chaotic systems via two examples. we first consider the fractional
order Chua's system [9]
\begin{equation}
\begin{array}{l}
\frac{d^\alpha x}{dt^\alpha} =a\left[y+\frac{x-2x^3}{7}\right]\\
\frac{d^\alpha y}{dt^\alpha}=x-y+z\\
\frac{d^\alpha z}{dt^\alpha}=-\frac{100}{7}y
\end{array}
\end{equation}
when $\alpha\geq 0.9$, this system can produce chaotic solutions
[9]. Particularly, when $\alpha=0.9$ and $a=12.75$, the fractional
order Chua's system is chaotic. The phase plot of $x$ and $z$ is
shown in Fig.1.
\begin{figure}[htb]
\centering
\includegraphics[width=6cm]{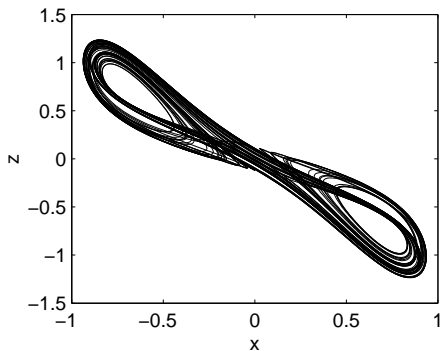}
\caption{Phase plot of the fractional order Chua's system with
$\alpha=0.9$.}
\end{figure}
\begin{figure}[htb]
\centering
\includegraphics[width=8.5cm]{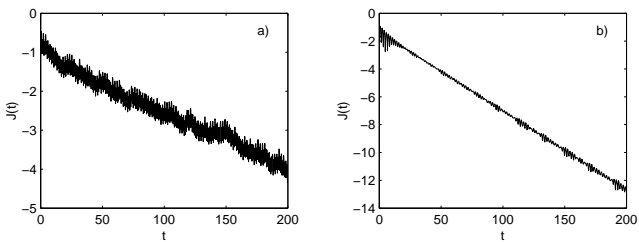}
\caption{Synchronization error of the fractional order Chua's
systems with $\alpha=0.9$: (a) $c=4$, (b) $c=7$.}
\end{figure}
\begin{figure}[htb]
\centering
\includegraphics[width=8.5cm]{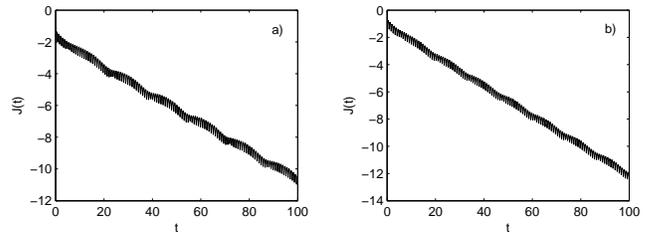}
\caption{Synchronization error of the integer order Chua's
systems: (a) $c=4$, (b) $c=7$.}
\end{figure}

We let $\Gamma=\mbox{diag}(1,0,0)$, which implies that only the
first variable $x$ is used to coupling the two fractional order
chaotic systems. To obtain a critical value of $c$ to make the two
systems synchronized, we continuously increase the coupling
strength $c$, from $c=0$, in step 0.5. When $c<4$, no synchronous
phenomenon is observed. When $c=4$, the curve of the
synchronization error $J(t)=\mbox{log}(\|e(t)\|)$ is shown in
Fig.2 (a), which indicate that the master-slave synchronization is
achieved. In Fig.2 (b), we show the curve of the synchronization
error when $c=7$, in which the synchronization effect is better
than that of $c=4$.

For the purpose of comparison, we also plot the curves of
synchronization error of the integer order Chua's systems
($a=9.5$) in Fig.3. Comparing Fig. 2 with Fig.3, we can know that
the synchronization rate of the fractional order Chua's systems is
slower than its integer order counterpart.

We next consider the fractional order R\"ossler system [16]:
\begin{equation}
\begin{array}{l}
\frac{d^\alpha x}{dt^\alpha} =-(y+z)\\
\frac{d^\alpha y}{dt^\alpha}=x+ay\\
\frac{d^\alpha z}{dt^\alpha}=0.2+z(x-10)
\end{array}
\end{equation}
when $\alpha=0.9$ and $a=0.4$, the above system is chaotic. The
phase diagram of the chaotic attractor is shown in Fig. 4.

We also let $\Gamma=\mbox{diag}(1,0,0)$, and do the similar
simulations as in the above example. When the coupling strength
$c=0.5$, the two fractional R\"ossler systems achieve
synchronization. The curve of the synchronization error of the
fractional order R\"ossler system is shown in Fig. 5 (a). In Fig.
5 (b), we also plot the curve of the synchronization error of the
integer order R\"ossler systems ($a=0.165$). From Fig. 5, we know
that the synchronization rate of the fractional order R\"ossler
systems is also slightly slower than its integer counterpart.
\begin{figure}[htb]
\centering
\includegraphics[width=7cm]{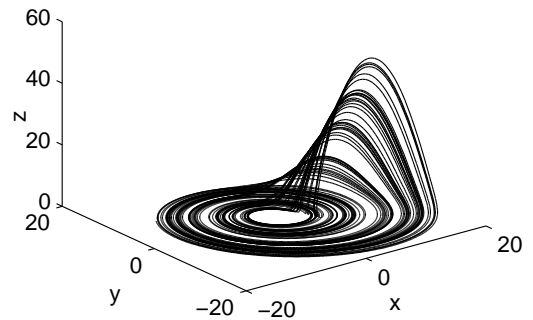}
\caption{Phase plot of the fractional order R\"ossler system with
$\alpha=0.9$.}
\end{figure}
\begin{figure}[htb]
\centering
\includegraphics[width=8.5cm]{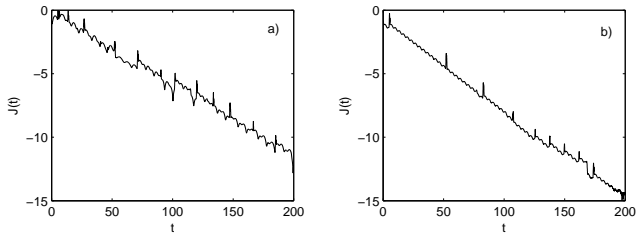}
\caption{The curves of synchronization error: (a) the fractional
order R\"ossler systems with $\alpha=0.9$ and $c=0.5$; (b) the
integer order R\"ossler systems with $c=0.5$.}
\end{figure}

We have also tested the synchronization scheme (2) on several
other fractional order chaotic systems [23]. Limited to the length
of this brief report, we omit these results here.

In summary, in this brief report, we have studied the master-slave
synchronization of fractional order chaotic systems. To our best
knowledge, this is the first report on the synchronization of
fractional order dynamical systems. We have shown that fractional
order chaotic systems can be synchronized by utilizing the similar
scheme as that of their integer order counterparts.

Future works regarding this topic include the investigation of
some other types of synchronization of fractional order chaotic
systems, such as the phase synchronization [24] and the projective
synchronization [25], as well as the synchronization of fractional
order hyperchaotic systems [16].

We acknowledge supports from the National Natural Science
Foundation of China under Grant 60271019, and the Youth Science
and Technology Foundation of UESTC under Grant YF020207.


\begin{thebibliography}{20}
\bibitem{"[13]"}
I. Grigorenko, and E. Grigorenko, Phys. Rev. Lett. 91, 034101
(2003). But the Eq. (5) in this paper is not correct, so the
results presented in this paper are not reliable, which we have
pointed out to the first author of this paper via personal
communication.
\end{thebibliography}
\end{document}